\documentclass[reprint,prd,twocolumn,superscriptaddress,showpacs,showkeys,floatfix,preprintnumbers]{revtex4-2}
\pdfoutput=1
\usepackage{mathrsfs}
\usepackage{amsfonts}
\usepackage{amsmath}
\usepackage{slashed}
\usepackage{array}
\usepackage{verbatim}
\usepackage{epsfig}
\usepackage{graphicx}
\usepackage{color}
\usepackage{subfigure}
\usepackage{multirow}
\usepackage[dvipsnames]{xcolor}
\definecolor{ceruleanblue}{rgb}{0.16, 0.32, 0.75}
\usepackage[colorlinks=true,linkcolor=ceruleanblue,citecolor=ceruleanblue,urlcolor=blue]{hyperref}
\usepackage{ulem}

\newcommand{\beq}{\begin{eqnarray}}
\newcommand{\eeq}{\end{eqnarray}}
\newcommand{\be}{\begin{equation}}
\newcommand{\ee}{\end{equation}}
\newcommand{\ba}{\begin{eqnarray}}
\newcommand{\ea}{\end{eqnarray}}
\newcommand{\nn}{\nonumber}
\newcommand{\bit}{\begin{itemize}}
\newcommand{\eit}{\end{itemize}}
\newcommand{\rw}{\rightarrow}

\newcommand{\zzuphy}{School of Physics, Zhengzhou University, Zhengzhou, Henan 450001, China}
\newcommand{\innovation}{Collaborative Innovation Center of Quantum Matter, Beijing 100871, China}
\newcommand{\chep}{Center for High Energy Physics, Peking University, Beijing 100871, China}
\newcommand{\pkuphy}{School of Physics, Peking University, Beijing 100871,
China}
\newcommand{\ccnu}{Key Laboratory of Quark \& Lepton Physics (MOE) and Institute of Particle Physics, Central China Normal University, Wuhan 430079, China}

\newcommand{\xian}{School of Sciences, Xi’an Technological University, Xi’an 710021, China}

\begin{document}

\title{Lattice determination of the neutrino background for $J/\psi \rightarrow \gamma + \textrm{invisible}$}

\author{Yu Meng}
\email[Email: ]{yu\_meng@zzu.edu.cn}
\affiliation{\zzuphy}
\author{Ning Li}\affiliation{\xian}
\author{Chuan Liu}
\email[Email: ]{liuchuan@pku.edu.cn}
\affiliation{\pkuphy}\affiliation{\chep}\affiliation{\innovation}
\author{Haobo Yan}\affiliation{\pkuphy}
\author{Ke-Long Zhang}\affiliation{\ccnu}
\author{Xue-Ze Zhang}\affiliation{\zzuphy}

\date{\today}

\begin{abstract}
Searching for dark matter is a primary goal of modern astronomy and particle physics. Invisible decays of heavy quarkonia are particularly promising for probing light dark matter, attracting broad interest due to their unique sensitivity. Experiments searching for radiative invisible decays of the $J/\psi$ have steadily improved upper limits, and upcoming facilities will push sensitivity further—making the precise determination and subtraction of the neutrino background indispensable. Here, we present the first lattice QCD calculation of the Standard Model decay $J/\psi \to \gamma\nu\bar{\nu}$, an irreducible background to $J/\psi \to \gamma + \textrm{invisible}$. Our result for the branching fraction is $\operatorname{Br}(J/\psi \to \gamma\nu\bar{\nu})=1.04(7)(8)\times 10^{-10}$, where the first uncertainty is statistical and the second is our systematic estimate. This work advances lattice-based determinations of neutrino backgrounds to quarkonium invisible decays, delivering an ab initio benchmark for $J/\psi \to \gamma + \textrm{invisible}$. Our approach generalizes to other quarkonium channels (e.g., $\Upsilon/\phi \to \gamma+\textrm{invisible}$) and provides critical theoretical support for dark matter searches at colliders.
\end{abstract}

\maketitle

\section{Introduction}
Over the past decades, abundant experimental observations have clearly hinted at the existence of dark matter, which triggered significant theoretical efforts to understand its nature and search for new physics beyond the Standard Model~\cite{Bertone:2004pz,Arkani-Hamed:2008hhe,Cadamuro:2011fd,Graham:2015ouw,Bauer:2021mvw}. Heavy-quarkonium systems offer a particularly clean laboratory for exploring dark-sector couplings to heavy quarks: unlike nuclear-recoil experiments operating at milli-eV scales, the decay of a heavy quark-antiquark bound state into a single photon plus missing energy is sensitive to arbitrarily light dark-sector particles. Consequently, radiative quarkonium transitions are now one of the standard candles in searches for light sterile neutrinos or sub-GeV dark matter, and has attracted extensive attention from numerous experimental and theoretical studies~\cite{Edwards:1982zn,Druzhinin:1987nx,CLEO:2010juh,BaBar:2010eww,Belle:2018pzt,BESIII:2020sdo,BESIII:2022rzz,BESIII:2021ges,Gao:2014yga,Li:2021phq,Colquhoun:2025xlx}.

Searches for new particles via invisible quarkonium decays have been ongoing for more than four decades. In 1982, the Crystal Ball experiment used $J/\psi$ radiative decay to look for a possible axion, setting an upper limit on the branching ratio $\operatorname{Br}(J/\psi\rw \gamma + \textrm{axion}) < 1.4\times 10^{-5}$ with the axion mass below 1 GeV~\cite{Edwards:1982zn}. Shortly thereafter, the VEPP-2M experiment placed a limit of $0.7\times 10^{-5}$ using $\phi$ radiative decay~\cite{Druzhinin:1987nx}. Since the beginning of this century, a succession of experiments with higher sensitivity have joined the hunt.
The CLEO~\cite{CLEO:2010juh}, BaBar~\cite{BaBar:2010eww}, Belle~\cite{Belle:2018pzt}, and BESIII~\cite{BESIII:2020sdo} experiments have performed the searches for $J/\psi$ or $\Upsilon$ radiative decays into invisible particles. Although no signal has been observed, the upper limits on the branching fractions have been improved significantly. The latest upper limits up to $10^{-8}$ on the branching fraction of $J/\psi \rw \gamma+\textrm{invisible}$ is reported by the BESIII experiment using $(2708.1\pm 14.5)\times 10^{6} \psi(3686)$ events collected by the detector~\cite{BESIII:2022rzz}, where the most stringent constraints on the ALP-photon coupling are given. Besides, a search for CP-odd light Higgs boson ($A^0$) in $J/\psi \rw \gamma A^0$ is also presented~\cite{BESIII:2021ges}. Among these searches, the neutrino background from the decay of $J/\psi \rw \gamma\nu\bar{\nu}$ is evidently contaminated since the neutrinos are also invisible particles in the Standard Model. Therefore, the contribution of this background component must be determined first, so that the experiment can remove it before searching for invisible matter. Finally, the remaining contribution will then truly originate from the invisible sector. 

At present, several futural experiments, such as Super Tau Charm Facility(STCF)~\cite{Achasov:2023gey,Ai:2025xop}, Belle II~\cite{Belle-II:2018jsg}, LHCb~\cite{LHCb:2018roe}, and CEPC~\cite{CEPCStudyGroup:2025kmw}, have the great potential to significantly improve the upper limit on the branching fraction of $J/\psi \rw \gamma+\textrm{invisible}$. In particular, the STCF is designed to accumulate $3.4\times 10^{12}$ $J/\psi$ events per year~\cite{Lyu:2021tlb}. Over five years of operation, this will produce a sample $10^5$ times larger than the $J/\psi$ statistics used by BESIII in their current invisible-decay analyses~\cite{BESIII:2022rzz}, which set an upper limit of $10^{-7}$. The STCF is therefore expected to push the sensitivity for $J/\psi\rightarrow \gamma + \textrm{invisible}$ down to the $10^{-10}$ level, entering the neutrino-background floor entirely. Hence,  at the present stage, an accurate prediction of the neutrino-background for $J/\psi \rightarrow \gamma + \textrm{invisible}$ is indispensable. 

 In this paper, we present the first lattice calculation of the invisible decay $J/\psi \rw \gamma\nu\bar{\nu}$. A genuine non-perturbative calculation can not only yield a model-independent prediction but also offers essential theoretical support for experimental searches for new physics beyond the Standard Model. The aim of the work is to non-perturbatively determine the branching fraction, with various systematic effects carefully examined.
To this end, the following strategies are adopted:
i) We use a method to extract the form factors with full $q^2$ distribution in the phase space.
ii) We eliminate the excited-state contamination via a multi-state fit, which is found to be non-negligible in this study.
iii) We perform a spatial volume integral
to obtain the decay width with a truncation range introduced to monitor the finite-volume effects.
iv) We utilize three ensembles with three different lattice spacings to perform a continuum extrapolation with the discretization effect well controlled. Our result will enable experiments to push the sensitivity for new physics searches in the $J/\psi \rightarrow \gamma + \textrm{invisible}$ channel down to the $10^{-10}$ level.

The rest of this paper is organized as follows. In Sec.~\ref{sec:method}, we
introduce the method for calculating the decay width on the lattice in this work. In Sec.~\ref{sec:setup} we give details of the simulations. In Sec.~\ref{sec:result} the numerical results of the decay width in all ensembles are obtained, a continuum extrapolation under three lattice spacings is performed, and the final result with the systematic error included is reported; Finally, we conclude in Sec.~\ref{sec:conclusion}.

\section{Methodology}\label{sec:method}
\begin{figure}[!h]
\centering
\subfigure{\includegraphics[width=0.35\textwidth]{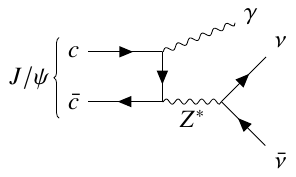}}\hspace{5mm}
\caption{An example of the feynman diagram for the decay $J/\psi\rw \gamma\nu\bar{\nu}$ with the photon emitted from the charm quark.}
\label{fig:diagram}
\end{figure}

For the decay $J/\psi\rw \gamma \nu\bar{\nu}$, the $J/\psi$ particle first emits a photon, which can be radiated by either the charm or the anticharm quark. The $\bar{c}c$ pair then couples to a virtual $Z$ boson, which subsequently decays into a neutrino–antineutrino pair. The Feynman diagram for photon radiation from the charm quark is shown in Fig.~\ref{fig:diagram}, while the diagram with the photon radiated from the anticharm quark is analogous and is omitted here. As hadrons are bound by the strong interaction, their internal dynamics are inherently non-perturbative. However, in electromagnetic transitions where a hadron radiates a photon, or in weak decays involving leptons and neutrinos, the coupling between the hadron and the photon or the weak gauge boson remains small and can be treated perturbatively. Consequently, the decay amplitude can be factorised into a product of a non-perturbative strong-interaction part and a perturbative leptonic part (whenever leptons participate).
As far as the amplitude of $J/\psi\rw \gamma\nu\bar{\nu}$ is concerned, the lowest-order contribution of which is expressed by
\beq\label{eq:amp}
\mathcal{M}&=&H_{\mu\nu\alpha}(q,p)\epsilon^{\alpha}_{J/\psi}(p)(-ie\epsilon^{\nu*}(q))(-\frac{i}{2}g_Z)^2 \nn \\
&\times&
\bar{u}(q_1)\frac{\gamma^{\mu}}{2}(1-\gamma_5)v(q_2)\frac{-i}{(k^2-m_Z^2)}
\eeq
where the nonperturbative hadronic interaction between the $J/\psi$, photon and $Z$ boson is encoded in a hadronic function $H_{\mu\nu\alpha}(q,p)$,
\beq\label{eq:hadron_mom}
H_{\mu\nu\alpha}(q,p)=\int d^4xe^{iqx}\mathcal{H}_{\mu\nu\alpha}(x,p)
\eeq
where the hadronic function $\mathcal{H}_{\mu\nu\alpha}(x,p)$ is defined as
\be\label{eq:hadron_function_x}
\mathcal{H}_{\mu\nu\alpha}(x,p)=\langle 0|T\{J_{\mu}^{\textrm{em}}(x)J_{\nu}^{Z}(0)\}|J/\psi(p)_{\alpha} \rangle
\ee
with the $J/\psi$ four-momentum $p=(m_{J/\psi},\vec{0})$, photon $q=(|\vec{q}|,\vec{q})$ and neutrino four-momenta $q_i=(|\vec{q}_i|,\vec{q}_i)$,$i=1,2$, both the photons and neutrinos satisfy the on-shell conditions and are treated as massless. The electromagnetic and weak currents are defined as $J_\mu^{\textrm{em}}=\sum_qe_q\,\bar{q}\gamma_\mu q$(with $e_q=2/3,-1/3,-1/3,2/3$ for $q=u,d,s,c$) and $J_\nu^{Z}=\sum_q \bar{q}\gamma_\nu(g_V^q-g_A^q\gamma_5) q$, where $g_V^{q}=T_3^{q}-2e_q\sin^2\theta_W$ and $g_A^q=T_3^q$, and $T_3^q$ denotes the third component of the weak isospin for the fermion. $\epsilon_{J/\psi}^{\alpha}(p)$ is the polarization vector of $J/\psi$, and $\epsilon^{\nu}(q)$ that of the photon. $e$ is the electromagnetic coupling constant, and $g_Z$ the $Z$-boson-fermion coupling constant. The $Z$-boson mass is $m_Z$, with its four-momentum given by $k=q_1+q_2$.

For the virtual $Z$-boson, $k^2 \ll m_Z^2 $, and it is natural to make a replacement for the $Z$-boson propagator
\be
\frac{1}{k^2-m_Z^2}\rw -\frac{1}{m_Z^2}
\ee

We also adopt the following notations:
\be
\frac{G_F}{\sqrt{2}}=\frac{g_W^2}{8m_W^2}, g_Z=\frac{g_W}{\cos\theta_W}, \cos\theta_W=\frac{m_W}{m_Z}
\ee
where $G_F$ is the Fermi coupling constant. The amplitude in Eq.~(\ref{eq:amp}) is thereby reduced to
\beq\label{eq:amp2}
\mathcal{M}&=&-e\frac{G_F}{\sqrt{2}}H_{\mu\nu\alpha}(q,p)\epsilon^{\alpha}_{J/\psi}(p)\epsilon^{\nu*}(q)\nn \\
&\times&\bar{u}(q_1)\gamma^{\mu}(1-\gamma_5)v(q_2)
\eeq
With consideration of the gauge symmetry and parity, the hadronic function $H_{\mu\nu\alpha}(q,p)$ can be parameterized as~\cite{Gao:2014yga}
\be\label{eq:form_factor}
H_{\mu\nu\alpha}(q,p)\equiv \epsilon_{\mu\nu\alpha\beta}q_{\beta}F_{\gamma\nu\bar{\nu}}(q^2)
\ee
In the rest frame of the $J/\psi$, a direct calculation of the decay width for $J/\psi \rw \gamma\nu\bar{\nu}$ using Eqs.~(\ref{eq:amp2}) and (\ref{eq:form_factor}) yields
\beq\label{eq:width}
&&\Gamma(J/\psi\rw \gamma\nu\bar{\nu}) \nn \\
&=&\frac{1}{2m_{J/\psi}}\int\frac{d^3\vec{q}}{(2\pi)^32|\vec{q}|}\int\frac{d^3\vec{q_1}}{(2\pi)^32|\vec{q}_1|}\int\frac{d^3\vec{q}_2}{(2\pi)^32|\vec{q}_2|} \nn \\
&\times&(2\pi)^4\delta^4(p-q-q_1-q_2)\times \frac{1}{3}|\mathcal{M}|^2 \times 3 \nn \\
&=&\frac{\alpha G_{F}^2}{3\pi^2}\int_{0}^{\frac{m_{J/\psi}}{2}} |\vec{q}|^3(m_{J/\psi}-|\vec{q}|)|F_{\gamma\nu\bar{\nu}}|^2 d|\vec{q}|
\eeq
where $\alpha\equiv e^2/4\pi$. The factor 1/3 in the third line denotes the average over the three polarizations of the $J/\psi$ in its rest frame, and the factor of 3 accounts for the three neutrino flavors.

The hadronic function $\mathcal{H}_{\mu\nu\alpha}(x,p)$ defined in Eq.(\ref{eq:hadron_function_x}) can be extracted from the Euclidean three-point function $C_{\mu\nu\alpha}^{(3)}(x;\Delta t)$ as 
\be
C_{\mu\nu\alpha}^{(3)}(x;\Delta t)=
\left\{
\begin{array}{lr}
\langle J_{\mu}^{\textrm{em}}(x)J_{\nu}^{Z}(0)\phi^{\dagger}_{J/\psi,\alpha}(-\Delta t) \rangle, &t\geq 0 \\
\langle J_{\mu}^{Z}(0) J_{\nu}^{\textrm{em}}(x)\phi^{\dagger}_{J/\psi,\alpha}(t-\Delta t) \rangle, &t<0 \\
\end{array}
\right.
\ee
by Wick rotation. A detailed discussion of this issue is summarized in the Appendix.~\ref{sec:app_A}. $\phi_{J/\psi,\alpha}=\bar{c}\gamma_{\alpha}c$ is the $J/\psi$ interpolating operator. A sufficiently large $\Delta t$ should be chosen to guarantee $J/\psi$ ground-state dominance. For a finite $\Delta t$, the hadronic function has a $\Delta t$ dependence, so we denote the hadronic function $\mathcal{H}_{\mu\nu\alpha}(x,p)$ as $\mathcal{H}_{\mu\nu\alpha}(x,\Delta t)$. For this hadronic function, the initial momentum $p$ is omitted since our calculation is restricted to the rest frame. We thus have
\be\label{eq:3pt_lat}
\mathcal{H}_{\mu\nu\alpha}(x,\Delta t)=
\left\{
\begin{array}{lr}
\frac{2m_{J/\psi}}{Z_0}e^{m_{J/\psi}\Delta t} C^{(3)}_{\mu\nu\alpha}(x;\Delta t), &t\geq 0 \\
\frac{2m_{J/\psi}}{Z_0}e^{m_{J/\psi}(\Delta t-t)} C^{(3)}_{\mu\nu\alpha}(x;\Delta t), &t<0 \\
\end{array}
\right.
\ee
with $Z_0=\langle J/\psi | \phi_{J/\psi}^{\dagger}|0\rangle$ the overlap amplitude for the $J/\psi$ ground state. Both $Z_0$ and $m_{J/\psi}$ can be calculated from the two-point function $C^{(2)}(t)=\langle \phi_{J/\psi}(t)
\phi^{\dagger}_{J/\psi}(0)\rangle$, which has the following expression:
\be\label{eq:2pt}
C^{(2)}(t)=\sum_{i=0,1}\frac{Z_i^2}{2E_i} \left(e^{-E_it}+e^{-E_i(T-t)}\right)
\ee
We adopt a two-state fit form for $C^{(2)}(t)$ to extract $Z_i$ and $E_i(i=1,2)$, where $E_0=m_{J/\psi}$ is the ground-state energy, $E_1$ the energy of the first excited state, and $Z_1$ the overlap amplitude for the first excited state.

To compute $F_{\gamma\nu\bar{\nu}}$, the traditional approach is to choose a series of lattice momenta $\vec{q}=2\pi\vec{n}/L$ with $\vec{n}=[0\ 0\ 1],[0\ 1\ 1],[1\ 1\ 1], \cdots$, and the phase-space integral is ultimately completed by interpolating or fitting this discrete $F_{\gamma\nu\bar{\nu}}(|\vec{q}|)$, leading to a model-dependent systematic uncertainty. In this work, we adopt an alternative approach, widely referred to as the scalar function method. The method has been extensively
applied to charmed systems~\cite{Meng:2021ecs,Zou:2021mgf,Meng:2024gpd,Meng:2024nyo,Meng:2024axn,Fan:2025qgj},  achieving remarkable success in precise calculations.

Following the parameterization of $H_{\mu\nu\alpha}(q,p)$ in Eq.~(\ref{eq:form_factor}), we construct the scalar function $\mathcal{I}$ by multiplying $\epsilon_{\mu\nu\alpha\beta}p_{\beta}$ to both sides. After averaging over the direction of $\vec{q}$, we obtain
\beq\label{eq:I_function}
&&\mathcal{I}(E_{\gamma},\Delta t) \nonumber \\
&&=im_{J/\psi}\int e^{E_{\gamma}t}dt\int d\vec{x}j_0(E_{\gamma}|\vec{x}|)\epsilon_{\mu\nu\alpha 0}\mathcal{H}_{\mu\nu\alpha}(x,\Delta t) \nonumber \\
\eeq
where $E_{\gamma}\equiv |\vec{q}|$. The form factor is then extracted as
\be
F_{\gamma\nu\bar{\nu}}(E_{\gamma},\Delta t)= -\frac{1}{6m_{J/\psi}E_{\gamma}}\mathcal{I}(E_{\gamma},\Delta t)
\ee

It should be emphasized that we use the hadronic function $\mathcal{H}_{\mu\nu\alpha}(\vec{x},\Delta t)$ calculated on a finite-volume lattice in Eq.~(\ref{eq:3pt_lat}) 
to replace the infinite-volume counterpart in Eq.~(\ref{eq:I_function}).
This replacement only gives rise to exponentially suppressed finite-volume effects, as $\mathcal{H}_{\mu\nu\alpha}(\vec{x},\Delta)$ itself is exponentially suppressed for large $|\vec{x}|$. A spatial integral cutoff $R$ can be introduced to check at large $R$ whether the finite-volume effects are well under control. Detailed checks are provided in the Appendix.~\ref{sec:app_FV}. As demonstrated in our previous studies on the charmonium system~\cite{Meng:2021ecs,Meng:2024nyo}, these finite-volume effects are negligible. 

Using $F_{\gamma\nu\bar{\nu}}(E_{\gamma},\Delta t)$ as input, the decay width for $J/\psi \rw \gamma\nu\bar{\nu}$ can be obtained via a Monte Carlo phase-integral over the region $E_{\gamma}\in [0,m_{J/\psi}/2]$:
\be\label{eq:decay_width_MC}
\Gamma_{\gamma\nu\bar{\nu}}(\Delta t)=\frac{\alpha G_{F}^2}{3\pi^2}\sum\limits_{i=1}^{N_{MC}}\left(E_{\gamma}^3(m_{J/\psi}-E_{\gamma})|F_{\gamma\nu\bar{\nu}}(E_{\gamma},\Delta t)|^2 \right)_i
\ee
where $N_{MC}$ is the number of Monte Carlo simulations, chosen such that the Monte-Carlo error is much smaller than the statistical error.

To further reduce lattice discretization effects, we define the dimensionless quantity $R_f\equiv \Gamma_{\gamma\nu\bar{\nu}}/f_{J/\psi}$, where $f_{J/\psi}$ is the decay constant of the $J/\psi$. For sufficiently large time $\Delta t$, $R_f(\Delta t)$ becomes independent of $\Delta t$, so a constant fit suffices to extract the final result. However, if $\Delta t$ is not large enough, such $\Delta t$ dependence can be parameterized using a simple two-state form:
\be\label{eq:th_fit}
R_f(\Delta t)=R_f+\zeta \cdot e^{-(E_1-E_0)\Delta t}
\ee
with two free parameters $R_f$ and $\zeta$. The difference between the two fitting results can be regarded as a systematic uncertainty associated with the excited-state contamination. After continuum extrapolation of both the dimensionless $R_f$ and the decay constant $f_{J/\psi}$, we obtain the physical results $R_f^{\textrm{Cont.Limit}}$ and $f_{J/\psi}^{\textrm{Cont.Limit}}$. The physical decay width is then obtained by rescaling $R_f^{\textrm{Cont.Limit}}$ with $f_{J/\psi}^{\textrm{phys}}$. Finally, the branching fraction is given by
\be
\operatorname{Br}[J/\psi \rw \gamma\nu\bar{\nu}]=R_f^{\textrm{Cont.Limit}}\times \frac{f_{J/\psi}^{\textrm{EXP}}}{\Gamma_{J/\psi}^{\textrm{EXP}}}
\ee
where $\Gamma_{J/\psi}^{\textrm{EXP}}=92.6$ keV is the world-average $J/\psi$ total decay width and $f_{J/\psi}^{\textrm{EXP}}$ is derived by experimental average of $\Gamma(J/\psi\rw e^+e^-)$ both from the Particle Data Group(PDG).

\section{Numerical setup}\label{sec:setup}

\begin{table}[!h]
\begin{ruledtabular}
\begin{tabular}{cccccc}
\textrm{Ensemble} & $a$ (fm) & $L^3\times T$ & $N_{\textrm{conf}}\times T$
& $m_{\pi} (\textrm{MeV})$ & $t$ \\
\hline
a67 & 0.0667(20) & $32^3\times 64$& $197\times 64$ & 300 & 12-18 \\
a85 & 0.085(2) & $24^3\times 48$ & $200\times 48$ & 315 & 10-14 \\
a98 & 0.098(3) & $24^3\times 48$ & $236\times 48$ & 365 & 9-13 \\
\end{tabular}
\end{ruledtabular}
\caption{
Parameters of gauge ensembles employed in this work. From left to right, we list the ensemble name, lattice spacing $a$, spatial and temporal lattice size $L$ and $T$, number of the correlation function measurements for each ensemble ($N_{\textrm{conf}}\times T$, with $N_{\textrm{conf}}$ the number of the gauge configurations), pion mass $m_{\pi}$, and the range of the time separation $t$ between the initial hadron and the electromagnetic current.
All quantities $L$, $T$, and $t$ are given in lattice units.}
\label{table:cfgs}
\end{table}

We use three two-flavor twisted-mass gauge ensembles generated by
the Extended Twisted Mass Collaboration (ETMC)~\cite{ETM:2009ptp,Becirevic:2012dc} with lattice spacings
$a \simeq 0.0667,0.085,0.098$ fm, respectively. For convenience, we denote them as a67, a85, and a98 in this work.
The ensemble parameters are shown in Table.~\ref{table:cfgs}. The valence charm quark mass is tuned
by setting the lattice-calculated $J/\psi$ mass to the physical value.
Details of this tuning are given in Ref.~\cite{Meng:2021ecs}.

In this work, we compute the three-point correlation function
$C^{(3)}_{\mu\nu\alpha}(\vec{x},t)$ using $Z_4$-stochastic wall source and point source. For time ordering $t\geq 0$, we place the point-source propagator on $J_{\nu}^{Z}$ and treat the electromagnetic current $J_{\mu}^{\textrm{em}}$ as the sink. For $t<0$, we exploit the space-time translation invariance of the correlation function, i.e.
$ \langle J_{\mu}^{Z}(0) J_{\nu}^{\textrm{em}}(x)\phi^{\dagger}_{J/\psi,\alpha}(t-\Delta t) \rangle =\langle J_{\mu}^{Z}(-\vec{x},-t) J_{\nu}^{\textrm{em}}(0)\phi^{\dagger}_{J/\psi,\alpha}(-\Delta t) \rangle$ and place the point-source propagator on $J_{\nu}^{\textrm{em}}$ with the weak current $J_{\mu}^{Z}$ as the sink. The wall-source propagator employed here reduces the mass spectrum uncertainty by nearly a factor of two. All propagators are generated on all time slices and averaged over to enhance the statistical precision, taking advantage of time translation invariance. We also apply APE~\cite{APE:1987ehd} and Gaussian smearing~\cite{Gusken:1989qx} to the $J/\psi$ interpolating field to effectively suppress excited-state effects.

In this work, we adopt the local vector electromagnetic current $J_\nu^{\textrm{em}}(x)=Z_V e_c \bar{c}\gamma_{\nu}c$ and weak $Z$-boson current $J_{\mu}^{Z}=\bar{c}\gamma_{\mu}(Z_Vg_V^c-Z_Ag_A^c\gamma_5)c$, where the renormalization constants $Z_V$ and $Z_A$ are introduced. A detailed determination of $Z_V$ was presented in our previous work~\cite{Meng:2021ecs}.
We directly quote these values here, which are 0.6047(19), 0.6257(21), and 0.6516(15) for $a=0.098,0.085,0.0667$ fm, respectively. The values of $Z_A$ are taken from Ref.~\cite{ETM:2010iwh}, where they are calculated via the RI-MOM scheme; the results are 0.746(11),0.746(06), and 0.772(06) for $a=0.098,0.085,0.0667$ fm, respectively.

\section{Numerical results}\label{sec:result}

\begin{figure}[!h]
\centering
\subfigure{\includegraphics[width=0.50\textwidth]{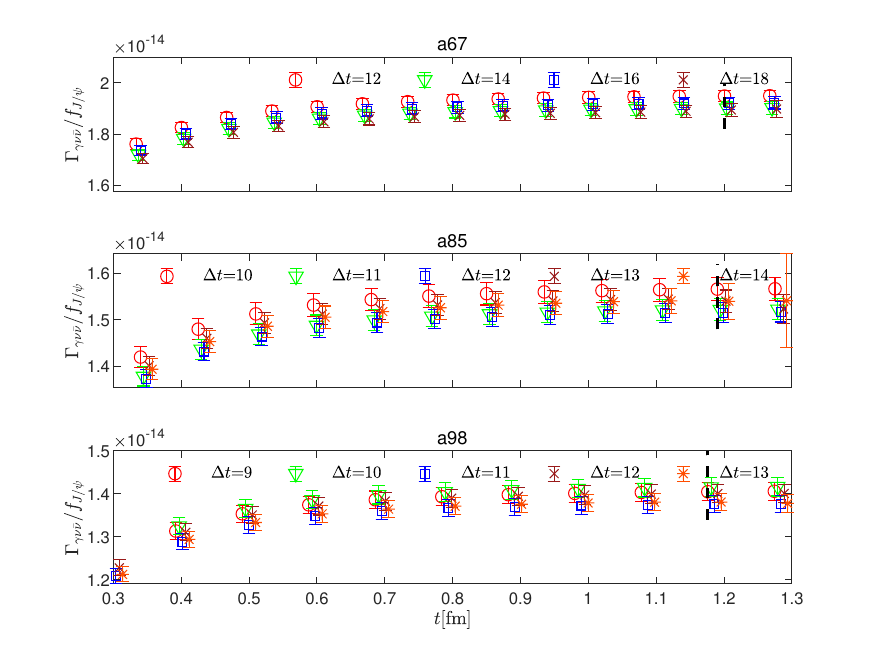}}\hspace{5mm}
\caption{The lattice results of $\Gamma_{\gamma\nu\bar{\nu}}/f_{J/\psi}$ for ensemble a67, a85 and a98, which are shown as a function of $t$ with various choices of $\Delta t$. The vertical dashed line denotes a conservative choice of $t \simeq 1.2$ fm, where the ground-state saturation is realized.}
\label{fig:F_th}
\end{figure}

\begin{figure}[!h]
\centering
\subfigure{\includegraphics[width=0.48\textwidth]{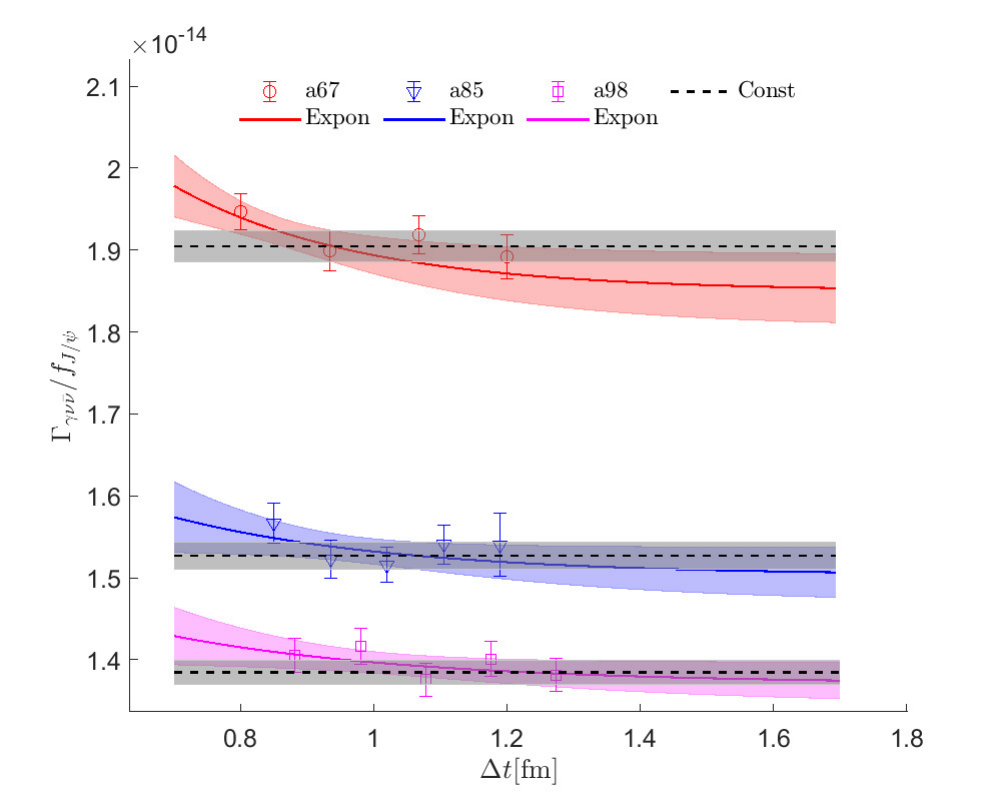}}\hspace{5mm}
\caption{The lattice results of $\Gamma_{\gamma\nu\bar{\nu}}/f_{J/\psi}$ with the cut $t\simeq 1.2$ fm in Fig.\ref{fig:F_th} are shown as a function of $\Delta t$ together with the constant fit and exponential form~(\ref{eq:th_fit}).}
\label{fig:F_th_limit}
\end{figure}

\begin{figure}[!h]
\centering
\subfigure{\includegraphics[width=0.48\textwidth]{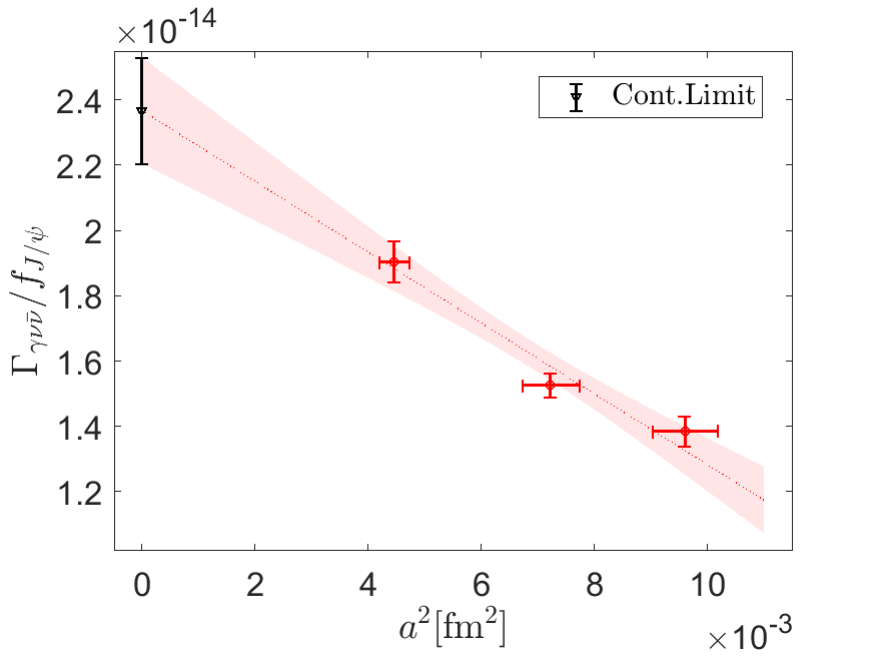}}\hspace{5mm}
\caption{Lattice values of $\Gamma_{\gamma\nu\bar{\nu}}/f_{J/\psi}$ as a function of lattice spacing together with a continuum extrapolation with a linear behavior $a^2$. The errors of lattice spacing are included in the fitting and presented by the horizontal error bars. The symbol of the red circle denotes the lattice results from ensemble a67,a85, and a98 from left to right. The statistical error of $Z_A$ is included by error propagation.}
\label{fig:width_cont_limit}
\end{figure}

The lattice results for $\Gamma_{\gamma\nu\bar{\nu}}/f_{J/\psi}$ as a function of $t$ for different separation $\Delta t$ are shown in Fig.~\ref{fig:F_th}. We find that for all values of $\Delta t$ and all ensembles employed in this work,
a temporal cutoff $t\simeq 1.2$ fm is a conservative choice to ensure ground-state saturation. With this choice, the results for $\Gamma_{\gamma\nu\bar{\nu}}/f_{J/\psi}$ as a function of $\Delta t$ are presented in Fig.~\ref{fig:F_th_limit}. At large time separations, a plateau spanning roughly three or four points emerges, which indicates that the ground state dominates in this regime. However, for small time separations, these results show a clear $\Delta t$ dependence of $\Gamma_{\gamma\nu\bar{\nu}}/f_{J/\psi}$, indicating non-negligible excited-state effects associated with $\phi_{J/\psi}^\dagger$ interpolating operator, as noted previously. To evaluate the systematic uncertainties arising from different fitting forms, we perform a constant fit at large time separations and the exponential fit incorporating all time separations as given by Eq.~(\ref{eq:th_fit}). They both enable us to extract the ground-state contribution to the ratio in the limit $\Delta t\to\infty$. The corresponding results are listed in Table~\ref{tab:width_ratio}. We adopt the constant-fit result as our central value, take the difference between the central values of the two fits as the systematic uncertainty, and obtain the total uncertainty by summing the two uncertainties in quadrature.

\begin{table}[!h]
\begin{ruledtabular}
\begin{tabular}{ccccc}
\textrm{Fitting form} & Exponential& Constant & Final  \\
\hline
a67 &  1.852(53) & 1.904(35)  & 1.904(63) \\
$\chi^2/\textrm{dof}$ & 0.88  & 0.50  \\
\hline
a85 & 1.503(42) & 1.527(29) & 1.527(38)   \\
$\chi^2/\textrm{dof}$ & 0.94  &  0.32 \\
\hline
98 & 1.371(48)  &  1.385(44) & 1.385(46) \\
$\chi^2/\textrm{dof}$ & 0.66  & 0.56  \\
\end{tabular}
\end{ruledtabular}
\caption{Numerical results of $\Gamma_{\gamma\nu\bar{\nu}}/f_{J/\psi}$ for three ensembles with different fitting forms.}
\label{tab:width_ratio}
\end{table}

In Fig.~\ref{fig:width_cont_limit}, the lattice results for $\Gamma_{\eta_c\gamma\gamma}/f_{J/\psi}$ at different lattice spacings are shown alongside a linear extrapolation in $a^2$. This linear behavior is expected due to the automatic $O(a)$ improvement inherent to twisted-mass configurations. The fitting curves are also found to describe the lattice data well. Following continuum extrapolation, we obtain $R_f^{\textrm{Cont.Limit}}=2.37(16)\times 10^{-14}$. For convenient future comparison with the experimental branching fraction, we convert $R_f^{\textrm{Cont.Limit}}$ to physical branching fraction by multiplying the $J/\psi$ decay constant $f_{J/\psi}^{\textrm{EXP}}=406.5(3.7)$ MeV, which is derived from the world average of $\Gamma(J/\psi\rw e^+e^-)=5.53(10)$ keV~\cite{ParticleDataGroup:2024cfk} and $\alpha_{QED}(m_{J/\psi}^2)=1/134.02(3)$~\cite{Hatton:2020qhk} via $\Gamma(J/\psi\rw e^+e^-)=4\pi/3\alpha^2_{QED}(m_{J/\psi}^2)e_c^2(f_{J/\psi}^{\textrm{EXP}})^2/m_{J/\psi}$. After dividing by the total decay width $\Gamma_{J/\psi}=92.6$ keV, the branching fraction is then obtained by $\operatorname{Br}[J/\psi \rw \gamma\nu\bar{\nu}]=1.04(7)\times 10^{-10}$. 

However, Ref.~\cite{ETM:2009ptp} note that the a98 ensemble may not be optimally tuned and could contain residual $\mathcal{O}(a)$ discretization errors. To investigate this effect, we also perform the continuum
extrapolation excluding the coarsest lattice ensemble, a98, yielding $1.12(11)\times 10^{-10}$. This result is consistent with $1.04(7)\times 10^{-10}$ but has a larger uncertainty. This consistency indicates the absence of residual $\mathcal{O}(a)$ effects for the $a98$ ensemble, a conclusion also supported by our recent work on charmonium systems~\cite{Meng:2021ecs,Zou:2021mgf,Meng:2024axn,Meng:2024czd} and
other lattice studies~\cite{ETM:2009ptp,Alexandrou:2009qu,ETM:2009ztk}.
We therefore quote the result including a98 as the final result, taking the difference between the two central values
as an estimation of the systematic uncertainty. Our final prediction for the branching fraction of $J/\psi \rw \gamma\nu\bar{\nu}$ is
\be
\operatorname{Br}[J/\psi \rw \gamma\nu\bar{\nu}]=1.04(7)(8)\times 10^{-10}
\ee
where the first error is a statistical error(including lattice spacing uncertainty from the extrapolation) and the second is our estimate of the systematic uncertainty.

A related phenomenological study based on the nonrelativistic color-singlet model predicts $\operatorname{Br}[J/\psi \rw \gamma\nu\bar{\nu}]=0.7\times 10^{-10}$~\cite{Gao:2014yga}. In contrast to above phenomenological model that only provide order-of-magnitude estimates with inherent approximations (e.g., nonrelativistic and color-singlet assumptions) that induce uncontrolled systematic uncertainties, our ab initio lattice calculation delivers a precise, parameter-free result free of such approximations. Our calculation uses three different lattice spacings for the continuum extrapolation, ensuring good control of the lattice discretization effects. We also employ multiple $\Delta t$ values and a multi-state fit to constrain the excited-state effects. Neglected disconnected diagrams are known to yield only small contributions
in charmonium system~\cite{McNeile:2004wu,deForcrand:2004ia,Levkova:2010ft,Hatton:2020qhk} due to Okubo-Zweig-Iizuka (OZI) suppression. Quenching the strange quark and employing up and down quarks heavier than their physical masses in our calculation are also known to introduce only minor effects. Limited by only three lattice spacings, the higher-order $a^4$ discretization effects remain unestimated. Nevertheless, such refinements can be more straightforwardly implemented in future lattice studies employing gauge ensembles with physical pion masses, heavy sea quarks, and a wider range of lattice spacings.

\section{Conclusion}\label{sec:conclusion}
In this paper, we present the first lattice QCD determination of the neutrino background in $J/\psi \rw \gamma+\textrm{invisible}$. Our calculation is performed using three $N_f=2$ twisted-mass fermion ensembles. Excited-state effects are observed and suppressed via a multi-state fit. After a controlled continuum extrapolation, we obtain a lattice QCD prediction for the branching fraction of $J/\psi \rw \gamma\nu\bar{\nu}$ as $\operatorname{Br}[J/\psi \rw \gamma\nu\bar{\nu}]=1.04(7)(8)\times 10^{-10}$. Here,  the first uncertainty is the statistical, already including the $a^2$-related uncertainty from the continuum extrapolation, and the second is an estimate of the systematic uncertainty.

Our ab initio calculation provides a precise prediction for the decay rate of $J/\psi \rw \gamma\nu\bar{\nu}$. Given that the future STCF is expected to reach a sensitivity of $10^{-10}$, the neutrino background must be precisely subtracted in the searches for new physics via the $J/\psi \rw \gamma+\textrm{invisible}$ channel. Furthermore, our study provides a crucial roadmap for searching for invisible decays in other quarkonium channels, such as $\Upsilon/\phi \rightarrow \gamma+\textrm{invisible}$. This work thus paves the way for further experimental and phenomenological investigations into potential dark matter signatures at particle colliders.

\begin{acknowledgments}
We thank ETM Collaboration for sharing the gauge configurations with us. We gratefully acknowledge the helpful discussions with Xu Feng, Dao-Neng Gao, and Xin-Yu Tuo.
The authors acknowledge support from NSFC under Grant No. 12305094, 12293060, 12293063. Y.M. also thanks the support from the Young Elite Scientists Sponsorship Program by Henan Association for Science and Technology with Grant No. 2025HYTP003. H.Y. acknowledges support from NSFC under Grant No. 124B2096. K.Z. thanks the support from the Cross Research Project of CCNU No. 30101250314. The numerical calculations are supported by the SongShan supercomputer at the National Supercomputing Center in Zhengzhou. The diagram is drawn using TikZ-Feynman~\cite{Ellis:2016jkw}.
\end{acknowledgments}

\bibliography{ref}
\bibliographystyle{apsrev4-2}

\appendix
\section{Relationship of hadronic function in Minkowski and Euclidean space}\label{sec:app_A}
In this section, we derive the relation between the hadronic functions in Minkowski and Euclidean spacetime, which is established by inserting a complete set of intermediate states into the respective hadronic functions.

In Minkowski spacetime, the hadronic function has the following decomposition
\begin{widetext}
\beq\label{eq:H_expand_M}
H_{\mu\nu\alpha}(q,p)
&=&i\sum\limits_{n,\vec{q}}\frac{1}{E_{\gamma}-E_{n}+i\epsilon}\langle 0|J_{\mu}^{\textrm{em}}(0)|n(\vec{q})\rangle \langle n(\vec{q})|J_{\nu}^{Z}(0)|J/\psi(p)_{\alpha} \rangle \nonumber \\
&-&i\sum\limits_{n',\vec{q}}\frac{1}{E_{\gamma}+E_{n'}-m_{J/\psi}-i\epsilon} \langle 0|J_{\nu}^{Z}(0)|n'(-\vec{q})\rangle \langle n'(-\vec{q})|J_{\mu}^{\textrm{em}}(0)|J/\psi(p)_{\alpha} \rangle \nonumber \\
\eeq
\end{widetext}
where the first line corresponds to the time-ordering $t>0$ and the second to $t<0$ in Eq.~(\ref{eq:hadron_mom}). The intermediate states $|n\rangle$ and $|n'\rangle$ represent all possible states with the allowed quantum numbers. For the connected contribution considered in this work, the low-lying states are given by $|n'\rangle=|J/\psi\rangle$ and $|n\rangle=|\eta_c\rangle$, respectively.

In the Euclidean spacetime, the hadronic function in Eq.~(\ref{eq:hadron_mom}) is replaced by $H_{\mu\nu\alpha}^E(q,p)$ obtained via a naive Wick rotation $t\rw -it$ as
\beq
H_{\mu\nu\alpha}^E(q,p)&=&-i\int_{-T/2}^{T/2} dt \int d^3\vec{x} e^{E_{\gamma}t-i\vec{q}\cdot \vec{x}}\mathcal{H}_{\mu\nu\alpha}(x,p) \nonumber \\
\eeq
with the Euclidean four-momenta $q=(iE_{\gamma},\vec{q})$ and $p=(im_{J/\psi},0)$.
As before, inserting a complete set of intermediate states into the Euclidean hadronic function above, we obtain
\begin{widetext}
\beq\label{eq:H_expand_E}
H_{\mu\nu\alpha}^E(q,p)
&=&i\sum\limits_{n,\vec{q}}\frac{1-e^{-(E_n-E_{\gamma})T/2}}{E_{\gamma}-E_{n}+i\epsilon}\langle 0|J_{\mu}^{\textrm{em}}(0)|n(\vec{q})\rangle \langle n(\vec{q})|J_{\nu}^{Z}(0)|J/\psi(p)_{\alpha} \rangle \nonumber \\
&-&i\sum\limits_{n',\vec{q}}\frac{1-e^{-(E_{\gamma}+E_{n'}-m_{J/\psi})T/2}}{E_{\gamma}+E_{n'}-m_{J/\psi}-i\epsilon} \langle 0|J_{\nu}^{Z}(0)|n'(-\vec{q})\rangle \langle n'(-\vec{q})|J_{\mu}^{\textrm{em}}(0)|J/\psi(p)_{\alpha} \rangle \nonumber \\
\eeq
\end{widetext}
where the finite time integral $[-T/2,T/2]$ is introduced to define the Euclidean hadronic function.

Whether the Minkowski hadronic function can be recovered from the Euclidean hadronic function via a naive Wick rotation generally depends on the convergence of all the $T$-dependent terms in the limit $T\rw \infty$. If these terms converge, the Wick rotation leaves the hadronic function unchanged, and the lattice calculation yield the physical results without particular difficulties. For this study, the conditions
\be\label{eq:condition_1}
E_n-E_{\gamma}>0
\ee
\be\label{eq:condition_2}
E_{\gamma}+E_{n'}-m_{J/\psi}>0
\ee
must be satisfied for $E_{\gamma} \in [0,m_{J/\psi}/2]$.

For the time ordering $t>0$, where the weak current is inserted before the electromagnetic current, the low-lying state is $J/\psi$ particle with momentum $\vec{q}$, and the condition (\ref{eq:condition_1}) is readily satisfied. However, the situation differs significantly for time ordering $t<0$, where the electromagnetic current is inserted before the weak current. In this case, the low-lying state is the $\eta_c$, whose mass is slightly less than that of the initial $J/\psi$ state. This results in a violation of condition (\ref{eq:condition_2}) for very small $E_{\gamma}$,e.g., $E_{\gamma}=0$. For all ensembles employed in this work, we find only the intermediate state $|\eta_c(\vec{q})\rangle$ with $\vec{q}=0$ violates condition (\ref{eq:condition_2}), leading to an exponentially growing factor $e^{-(E_{\gamma}+E_{n'}-m_{J/\psi})T/2}$ as $T$ increases. This can be verified numerically using the discrete energy levels of $\eta_c$ summarized in Table~\ref{tab:disper}. For $\vec{q}=0$, however, we have
\be
\langle 0|J_{\nu}^{Z}(0)|\eta_c(\vec{0})\rangle \langle \eta_c(\vec{0})|J_{\mu}^{\textrm{em}}(0)|J/\psi(\vec{0})_{\alpha} \rangle=0
\ee
which still protects the Euclidean hadronic function from the exponentially growing factor $e^{-(m_{\eta_c}-m_{J/\psi})T/2}$. In other words, all intermediate states with discrete momenta $\vec{q}=2\pi\vec{n}/L$ are independent of the $T$-dependence factor in the limit $T\rw \infty$. We conclude that for the time ordering $t<0$, condition (\ref{eq:condition_1}) is also satisfied in our calculations. Thus, we can extract the Minkowski hadronic function directly from the Euclidean hadronic function via a naive Wick rotation, and the $i\epsilon$ terms in Eq.(\ref{eq:H_expand_M}) and Eq.(\ref{eq:H_expand_E}) are unnecessary.

\begin{table}[!h]
\center
\begin{ruledtabular}
\begin{tabular}{cccccc}
Ensemble & a67 & a85 &a98 \\
$aE_{\eta_c}(|\vec{n}|^2=0)$ & 1.0142(2) & 1.2958(3) & 1.4995(3) \\
$aE_{\eta_c}(|\vec{n}|^2=1)$ & 1.0302(2) & 1.3157(3) & 1.5144(4) \\
$aE_{\eta_c}(|\vec{n}|^2=2)$ & 1.0467(2) & 1.3354(3) & 1.5290(4) \\
$aE_{\eta_c}(|\vec{n}|^2=3)$ & 1.0629(3) & 1.3546(4) & 1.5434(4) \\
$aE_{\eta_c}(|\vec{n}|^2=4)$ & 1.0782(4) & 1.3729(5) & 1.5572(5) \\
\hline
$a\delta E(|\vec{n}|^2=0)$ & -0.0343(2) & -0.0372(3) & -0.0387(3) \\
$a\delta E(|\vec{n}|^2=1)$ & 0.1781(2) & 0.2446(3) & 0.2380(4) \\
$a\delta E(|\vec{n}|^2=2)$ & 0.2758(3) & 0.3737(3) & 0.3611(4) \\
$a\delta E(|\vec{n}|^2=3)$ & 0.3544(3) & 0.4751(4) & 0.4587(4) \\
$a\delta E(|\vec{n}|^2=4)$ & 0.4223(4) & 0.5636(5) & 0.5426(5) \\
\end{tabular}
\end{ruledtabular}
\caption{Numerical results of $E_{\eta_c}(\vec{p})$ and $\delta E(\vec{p})$ with $\vec{p}=2\pi\vec{n}/L,|\vec{n}|^2=0,1,2,3,4$.}
\label{tab:disper}
\end{table}

It is also worth emphasizing that even if the Eq.~(\ref{eq:condition_2}) may not be satisfied in other calculations---particularly for larger lattice volumes---no particular obstacle arises in determining the decay width. This is because the mass of the intermediate $\eta_c$ is very close to that of the initial $J/\psi$, with the relative mass difference $\delta m_{cc}/m_{J/\psi}< 4\%$, where $\delta m_{cc}\equiv m_{J/\psi}-m_{\eta_c}$. Furthermore, with the current precision level of approximately $10\%$, the decay width is strongly suppressed by the $|\vec{q}|^3$ factor in Eq.~(\ref{eq:width}), rendering the effects from the exponential growth factors negligible within the current precision level.

\section{Finite-volume effects} \label{sec:app_FV}
The decay width is calculated by a Monte Carlo phase integral as shown in Eq.~(\ref{eq:decay_width_MC}), where we choose $N_{MC}=200$ and verify this choice to ensure the phase integral uncertainty is much smaller than the statistical uncertainty. In our calculations, the integral energy $E_{\gamma} \in [0,m_{J/\psi}/2]$ is sampled randomly.
Non-lattice values ($E_{\gamma}\neq 2\pi|\vec{n}|/L$) inevitably introduce systematic effects. These effects are essentially finite-volume effects, as all random $E_{\gamma}$ values reduce to lattice values in the limit $L\rightarrow \infty$. 

\begin{figure}[!h]
\centering
\subfigure{\includegraphics[width=0.48\textwidth]{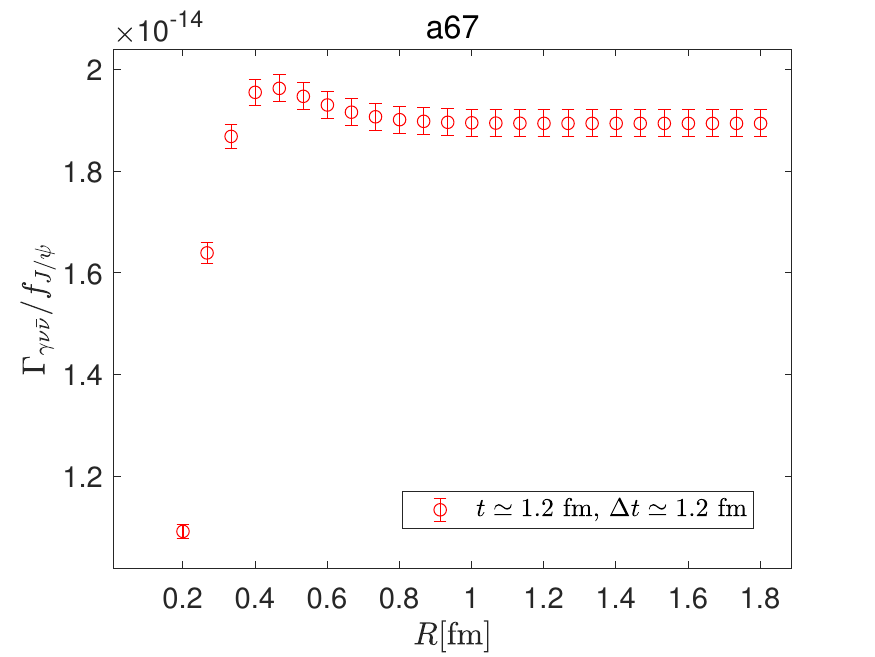}}\hspace{5mm}
\caption{For ensemble a67, $\Gamma_{\gamma\nu\bar{\nu}}/f_{J/\psi}$ with $t\simeq 1.2$ fm and $\Delta t \simeq 1.2$ fm as a function of the spatial range truncation $R$.}
\label{fig:R-dependence}
\end{figure}

To investigate finite-volume effects, we introduce a spatial integral truncation parameter $R$ in Eq.~(\ref{eq:I_function}).
Since the hadronic function $\mathcal{H}_{\mu\nu\alpha}(x)$ is dominated by the $\eta_c$ state for large $|\vec{x}|$, the integrand is exponentially suppressed as $|\vec{x}|$ increases.
In Fig.~\ref{fig:R-dependence}, the ratio $\Gamma_{\gamma\nu\bar{\nu}}/f_{J/\psi}$ is shown as a function of $R$. A clear plateau is observed for $R\gtrsim 0.8$ fm, indicating that the hadronic function $\mathcal{H}_{\mu\nu\alpha}(x)$
for $|\vec{x}|\gtrsim 0.8$ fm makes a negligible contribution to $\Gamma_{\gamma\nu\bar{\nu}}/f_{J/\psi}$.
All the ensembles have a lattice size $\frac{L}{2}>1$ fm which is sufficiently large to accommodate the hadron. We thus conclude that finite-volume effects are well under control in our calculations.

\end{document}